\documentclass[subeqn,12pt, a4paper]{article} 
\usepackage{amssymb,amsmath,amsfonts,amsthm, amscd, mathrsfs,helvet}
\usepackage{bm, stmaryrd}
\usepackage{graphicx,verbatim}
\usepackage[usenames, dvipsnames]{color}
\usepackage{psfrag}
\usepackage[all]{xy}
\usepackage{tikz}
\usetikzlibrary{arrows,matrix,decorations.markings,shapes.geometric}
\usepackage[normalem]{ulem}
\usepackage{xspace}

\theoremstyle{plain}

\newtheorem*{theorem*}{Theorem}

\newtheorem*{proposition*}{Proposition}

\newcommand{\tensor}[1]{{\bf \underline{#1}}}

\definecolor{brightBlue}{rgb}{0,0,1}

\definecolor{Violet}{rgb}{0.47,0,1}

\DeclareMathOperator{\tr}{tr}

\def\b{\mathfrak{b}}

\def\g{\mathfrak{g}}

\def\ha{\mbox{\small $\frac{1}{2}$}}

\newcounter{comcompt}
\newcounter{boxq}
\def\CC{\mathbb{C}}

\def\L{\mathcal{L}}

\def\1{\tensor{1}}
\def\2{\tensor{2}}
\def\3{\tensor{3}}
\def\4{\tensor{4}}

\def\beq{\begin{equation}}
\def\eeq{\end{equation}}
\def\beqz{\begin{equation*}}
\def\eeqz{\end{equation*}}
\def\bea{\begin{eqnarray}}
\def\eea{\end{eqnarray}}

\def\dss{\delta_{\sigma \sigma'}}
\def\pdss{\partial_\sigma \delta_{\sigma \sigma'}}
\def\KK{{\mathcal K}}
\newcommand{\pcm}{principal chiral model\xspace}

\newcommand{\pbs}{Poisson brackets\xspace}

\numberwithin{equation}{section}

\begin{document}

\begin{center}
\vspace*{2em}
{\large\bf Integrable double deformation of the principal chiral model}\\
\vspace{1.5em}
F. Delduc$\,{}^1$, M. Magro$\,{}^1$, B. Vicedo$\,{}^2$

\vspace{1em}
\begingroup\itshape
{\it 1) Laboratoire de Physique, ENS Lyon
et CNRS UMR 5672, Universit\'e de Lyon,}\\
{\it 46, all\'ee d'Italie, 69364 LYON Cedex 07, France}\\
\vspace{1em}
{\it 2) School of Physics, Astronomy and Mathematics,
University of Hertfordshire,}\\
{\it College Lane,
Hatfield AL10 9AB,
United Kingdom}
\par\endgroup
\vspace{1em}
\begingroup\ttfamily
Francois.Delduc@ens-lyon.fr, Marc.Magro@ens-lyon.fr, Benoit.Vicedo@gmail.com
\par\endgroup
\vspace{1.5em}
\end{center}

\paragraph{Abstract.}

We define a two-parameter family of integrable deformations of the principal chiral model on an arbitrary compact group. The Yang-Baxter $\sigma$-model and the principal chiral model with a Wess-Zumino term both correspond to limits in which one of the two parameters vanishes.

\section{Introduction}

It was shown by S. Rajeev in \cite{Rajeev:1988hq}  that there exists a
one-parameter deformation of the \pbs satisfied by  the current of the \pcm.
It was also observed that this deformed Poisson structure coincides with
that of two Poisson commuting classical Kac-Moody currents \cite{Rajeev:1988hq}.
It turns out that the Kac-Moody currents are either both real or
complex conjugate of one another, depending on the value of the
deformation parameter \cite{Hollowood:2014rla}. We will
therefore refer to these two branches in the deformation parameter
as real and complex, respectively.
In each branch, the integrable field theory which provides a realisation
of the deformed Poisson algebra is known.
  As was shown in \cite{Delduc:2013fga}, for the complex branch this is the Yang-Baxter $\sigma$-model defined by C. Klim\v{c}\'{\i}k in \cite{Klimcik:2002zj,Klimcik:2008eq}.
The Yang-Baxter $\sigma$-model on $SU(2)$ is simply the squashed 3-sphere $\sigma$-model
  studied in  \cite{Kawaguchi:2011pf,Kawaguchi:2012gp}. For the real branch, the model is the one
introduced by K. Sfetsos in \cite{Sfetsos:2013wia} (see also \cite{Hollowood:2014rla,Itsios:2014vfa,Sfetsos:2014cea}). It generalises the model   studied in \cite{Balog:1993es} to higher rank.

\medskip

In the case of $SU(2)$, it was found in \cite{Balog:1993es} that the deformed \pbs can even be extended to a two-parameter deformation. Furthermore, it is clear that the \pbs constructed in \cite{Balog:1993es}  extend immediately to any Lie algebra. In this note we exhibit the action of the integrable field theory which realises the double deformation in the complex branch.

\section{Action and Lax pair}

\subsection{Ansatz for the action}
Let $\g$ be a compact Lie algebra with Lie group $G$.
To construct a two-parameter deformation of the \pcm on $G$ we will start with a fairly general ansatz for the action of a $G$-valued field $g$, which we take to be:
\begin{align} \label{act-depart}
S[g] &= - \ha K \int d^2x \tr \Bigl(
g^{-1}\partial_- g \frac{1+\eta^2}{1-\eta^2 R^2} g^{-1}\partial_+ g
+ g^{-1}\partial_- g \frac{(1+\eta^2) A R}{1-\eta^2 R^2} g^{-1}\partial_+ g
 \Bigr) \notag \\
 &\qquad\qquad - \ha k K \int d^3 x \tr \Bigl(g^{-1} \partial_{\xi} g [g^{-1} \partial_- g , g^{-1} \partial_+ g] \Bigr).
\end{align}
Here $\partial_{\pm} = \partial_{\tau} \pm \partial_{\sigma}$ denote the usual light-cone derivatives on the worldsheet. The last term in \eqref{act-depart} is the standard Wess-Zumino term integrated over a 3-dimensional space parameterised by $(\tau, \sigma, \xi)$ and whose boundary is the worldsheet. We denote this term by $S_{\rm WZ}[g]$.

For the moment $\eta$, $A$ and $k$ are three independent real parameters. The real parameter $K$ is an overall normalisation which will not play much role in our analysis. Note that when $k=0$ and $A=\pm \eta$
one recovers the Yang-Baxter $\sigma$-model of \cite{Klimcik:2002zj,Klimcik:2008eq}.
Furthermore, when $\eta=A=k=0$ the above action reduces to that of the \pcm.

The linear operator $R$ is a skew-symmetric non-split solution of the modified classical Yang-Baxter equation (see
 \cite{Delduc:2013fga} for details) on $\g$. That is to say, for any $x, y \in \g$ it satisfies
 \begin{equation} \label{YB R}
[R x, R y] = R \big( [R x, y] + [x, R y] \big) + [x, y].
\end{equation}
In what follows we shall work with the standard solution to this equation which is constructed as follows. Given a Cartan-Weyl basis $(H^i,E^\alpha)$ of the complexified Lie algebra $\g^\CC$, a basis of the compact real form $\g$ is given by
\begin{equation*}
 T^i = i H^i, \quad B^\alpha = \frac{i}{\sqrt{2}} (E^{+\alpha} + E^{-\alpha}),
\quad C^\alpha =  \frac{1}{\sqrt{2}} (E^{+\alpha} - E^{-\alpha}).
\end{equation*}
The operator $R$ is then defined by \cite{Klimcik:2008eq}
\begin{equation} \label{august22m}
R(T^i)=0, \quad R(B^\alpha)=C^\alpha, \quad R(C^\alpha) = - B^\alpha.
\end{equation}

\subsection{Flat and conserved current}

The equations of motion computed from the action \eqref{act-depart} take the form of the
conservation equation
\beqz
\partial_- \KK_+ + \partial_+ \KK_- =0
\eeqz
where we have defined
\begin{equation} \label{KP-KM}
\KK_{\pm} = g \biggl( \Bigl( \frac{1+\eta^2}{1-\eta^2 R^2} (1 \pm AR) \mp k \Bigr) g^{-1}\partial_{\pm} g \biggr) g^{-1}.
\end{equation}
We will now determine the conditions under which this conserved current $\KK_\pm$ is also on-shell flat. This will immediately imply the existence of a Lax pair for the resulting model.
However, for simplicity, we shall make use of the fact that by construction the $R$-matrix \eqref{august22m} satisfies
\beq \label{R3 -R}
R^3 = - R.
\eeq
Using the property \eqref{R3 -R}  one can show that
\begin{equation} \label{inv op R}
\frac{1+\eta^2}{1 - \eta^2 R^2} = 1 + \eta^2 +  \eta^2  R^2, \qquad
\frac{(1+\eta^2)R}{1 - \eta^2 R^2} = R .
\end{equation}
The action \eqref{act-depart} may therefore be written as
\begin{equation} \label{action-ybwzw}
S[g]  = - \ha K \int d^2 x \tr \Bigl( g^{-1} \partial_- g (1+ \eta^2 + AR + \eta^2 R^2)
g^{-1}\partial_+ g  \Bigr) + S_{\rm{WZ}}[g].
\end{equation}

Substituting \eqref{inv op R} into the above expression \eqref{KP-KM} for $\KK_{\pm}$ we obtain
\begin{equation} \label{K as Igdg}
g^{-1} \KK_{\pm} g = \left( 1 + \eta^2 \mp k \pm A\, R + \eta^2 R^2 \right) g^{-1} \partial_{\pm} g.
\end{equation}
The inverse of the operator on the right hand side of \eqref{K as Igdg}
 can be constructed explicitly by equating the coefficients in front of each power of $R$, using \eqref{R3 -R}, on both sides of the following equation
\begin{equation*}
\left( a_{\pm} + b_{\pm} R + c_{\pm} R^2 \right) \left( 1 + \eta^2 \mp k \pm A\, R + \eta^2 R^2 \right) = 1.
\end{equation*}
We find explicitly that
\begin{equation} \label{abc def}
a_{\pm} = \frac{1}{1 + \eta^2 \mp k}, \qquad
b_{\pm} = \mp \frac{A}{A^2 + (1 \mp k)^2}, \qquad
c_{\pm} = \frac{1}{1 + \eta^2 \mp k} - \frac{1 \mp k}{A^2 + (1 \mp k)^2}.
\end{equation}
This inverse operator enables us to rewrite \eqref{K as Igdg} as
\begin{equation} \label{Igdg as J}
\partial_{\pm} g g^{-1} = \left( a_{\pm} + b_{\pm} R_g + c_{\pm} R_g^2 \right) \mathcal \KK_{\pm},
\end{equation}
where the action of the operator $R_g$ on any $x \in \g$ is defined as
\begin{equation} \label{def Rg}
R_g x = g R(g^{-1} x g) g^{-1}.
\end{equation}

Next we note that the modified classical Yang-Baxter equation
\eqref{YB R} for $R$ together with the property \eqref{R3 -R}
implies the following equation for $R^2$,
\begin{equation} \label{YB R2}
[R^2 x, R^2 y] = R^2 \big( [R^2 x, y] + [x, R^2 y] \big) + (1 + 2 R^2)[x, y].
\end{equation}
Substituting the expression \eqref{Igdg as J} into the curvature
\begin{equation*}
\partial_- \big( \partial_+ g g^{-1} \big) - \partial_+ \big( \partial_- g g^{-1} \big) + \big[ \partial_+ g g^{-1}, \partial_- g g^{-1} \big]
\end{equation*}
and making use of the modified classical Yang-Baxter equation \eqref{YB R} for the operator $R$ as well as its consequence \eqref{YB R2}, one can show that
\begin{align} \label{ZCeq}
\partial_- &\big( \partial_+ g g^{-1} \big) - \partial_+ \big( \partial_- g g^{-1} \big) + \big[ \partial_+ g g^{-1}, \partial_- g g^{-1} \big] \notag\\
&= \left( \frac{a_+ - a_-}{2} + \frac{b_+ - b_-}{2} R_g + \frac{c_+ - c_-}{2} R_g^2 \right) \left( \partial_- \KK_+ + \partial_+ \KK_- \right) \notag\\
&\qquad + \left( \frac{a_+ + a_-}{2} + \frac{b_+ + b_-}{2} R_g + \frac{c_+ + c_-}{2} R_g^2 \right) \left( \partial_- \KK_+ - \partial_+ \KK_- \right) \notag\\
&\qquad + \bigg( a_+ a_- - b_+ b_- - c_+ c_- + \bigr( (a_+ - c_+) b_- + (a_- - c_-) b_+ \bigl) R_g \notag\\
&\qquad\qquad\qquad\qquad + \bigr( (a_+ - c_+) c_- + (a_- - c_-) c_+ \bigl) R_g \bigg) [\KK_+, \KK_-] \notag\\
&\qquad + b_+ c_- (1+ R_g^2) [R_g \KK_-, \KK_+] + b_- c_+ (1+ R_g^2) [\KK_-, R_g \KK_+].
\end{align}

Now we have the following relations between the parameters \eqref{abc def},
\beq  \label{rel-coeff-1}
\frac{b_+ + b_-}{2} = (a_+ - c_+) b_- + (a_- - c_-) b_+.
\eeq
Moreover, we impose the relation
\beq \label{A-k-eta}
A = \eta \sqrt{1 - \frac{k^2}{1 + \eta^2}},
\eeq
between the three parameters $\eta$, $A$ and $k$. Note that the exact same relation \eqref{A-k-eta} was previously obtained in \cite{Kawaguchi:2011mz,Kawaguchi:2013gma} for the `squashed WZNW-model', which is a deformation of the squashed 3-sphere $\sigma$-model by addition of a WZ term. The model defined by the action \eqref{act-depart}, with $A$ set to the value \eqref{A-k-eta}, can therefore be seen as a generalisation of the latter from the $\mathfrak{su}(2)$ case to an arbitrary Lie algebra $\g$. Using \eqref{A-k-eta}, we find the following further relations among these parameters
\begin{subequations}\label{rel-coeff-2}
\begin{align}
b_+ c_- &= b_- c_+, \label{rel-coeff-2a} \\
\frac{a_+ + a_-}{2} &= a_+ a_- - b_+ b_- - c_+ c_-,\\
 \frac{c_+ + c_-}{2} &= (a_+ - c_+) c_- + (a_- - c_-) c_+. \label{rel-coeff-2b}
\end{align}
\end{subequations}
Substituting all these relations into \eqref{ZCeq} we arrive at the following
\begin{align} \label{ZCeq2}
\partial_- &\big( \partial_+ g g^{-1} \big) - \partial_+ \big( \partial_- g g^{-1} \big) + \big[ \partial_+ g g^{-1}, \partial_- g g^{-1} \big] \notag\\
&= \left( \frac{a_+ - a_-}{2} + \frac{b_+ - b_-}{2} R_g + \frac{c_+ - c_-}{2} R_g^2 \right) \left( \partial_- \KK_+ + \partial_+ \KK_- \right) \notag\\
&\qquad + \left( \frac{a_+ + a_-}{2} + \frac{b_+ + b_-}{2} R_g + \frac{c_+ + c_-}{2} R_g^2 \right) \left( \partial_- \KK_+ - \partial_+ \KK_- + [\KK_+, \KK_-] \right) \notag\\
&\qquad + b_+ c_- (1+ R_g^2) \bigr( [R_g \KK_-, \KK_+] + [\KK_-, R_g \KK_+] \bigl).
\end{align}
The last line can be shown to vanish using the following identity
\begin{equation} \label{id R2}
(1 + R^2) \big( [R x, y] + [x, R y] \big) = 0,
\end{equation}
valid for any $x, y \in \g$. 
To show the latter, note that the left hand side can be rewritten as
\begin{equation*}
(1+ R^2) \bigr( [R x, y] + [x, R y] \bigl)
= (R - i) \big[ (R + i) x, (R + i) y \big],
\end{equation*}
by using the modified classical Yang-Baxter equation \eqref{YB R} for $R$ in the form
\begin{equation*}
(R + i) \big( [R x, y] + [x, R y] \big) = \big[ (R + i) x, (R + i) y \big].
\end{equation*}
However, since $R \pm i$ are respectively the projectors onto the Borel subalgebras $\b_{\pm}$ of $\g^{\CC}$, it follows that $(R - i) \big[ (R + i) x, (R + i) y \big] = 0$ for any $x, y \in \g$, and hence we obtain \eqref{id R2}. We are therefore left with
\begin{align} \label{ZCeq3}
\partial_- &\big( \partial_+ g g^{-1} \big) - \partial_+ \big( \partial_- g g^{-1} \big) + \big[ \partial_+ g g^{-1}, \partial_- g g^{-1} \big] \notag\\
&= \left( \frac{a_+ - a_-}{2} + \frac{b_+ - b_-}{2} R_g + \frac{c_+ - c_-}{2} R_g^2 \right) \left( \partial_- \KK_+ + \partial_+ \KK_- \right) \notag\\
&\qquad + \left( \frac{a_+ + a_-}{2} + \frac{b_+ + b_-}{2} R_g + \frac{c_+ + c_-}{2} R_g^2 \right) \left( \partial_- \KK_+ - \partial_+ \KK_- + [\KK_+, \KK_-] \right).
\end{align}

Since the operator in parenthesis in the last line is invertible, this implies
that the conserved current $\KK_\pm$ is also on-shell flat. The equations
of motion for the action \eqref{act-depart} with the parameter $A$
fixed by the relation \eqref{A-k-eta} can therefore be recast in
the form of a zero curvature equation for the Lax pair
\begin{equation*}
\L_{\pm}(z) = \frac{\KK_{\pm}}{1\mp z}.
\end{equation*}
We denote the spatial component $\ha (\L_+(z) - \L_-(z))$ by
\begin{equation} \label{spatial lax}
\L(z) = \frac{1}{1-z^2} (\KK_1 + z \KK_0).
\end{equation}

\subsection{3-parameter current}

Before studying the 2-parameter action \eqref{act-depart} with $A$ fixed by \eqref{A-k-eta}, let us note that there is a redundancy between the set of four equations \eqref{rel-coeff-1} and \eqref{rel-coeff-2} for the six coefficients $(a_\pm,b_\pm,c_\pm)$ entering in the equation \eqref{Igdg as J}. Indeed, equations
\eqref{rel-coeff-1} and \eqref{rel-coeff-2a} together imply \eqref{rel-coeff-2b}. As a consequence, it is possible to construct a 3-parameter current\footnote{We thank B. Hoare for pointing out this
 possibility.} whose conservation equation implies its flatness. After solving the remaining
 independent equations, one may parameterize this
 current as
\begin{equation} \label{3-param current}
g^{-1} \KK_{\pm} g  =   \bigl(  1+ \eta^2 \mp   k  \pm A R + \eta^2 R^2 \pm
 \xi   R^2  \bigr) g^{-1}\partial_{\pm} g ,
\end{equation}
where $A$ is now given in terms of $k$, $\eta$ and $\xi$ as
\beqz
A = \eta \sqrt{1 - \frac{\frac{\xi^2}{\eta^2} + (k + \xi)^2}{1 + \eta^2}}.
\eeqz
 It is not clear, however, which action could give rise to such a current.

\section{Double deformation of the \pbs}

In this section, we will derive the Hamiltonian form of the fields
$\KK_0$, $\KK_1$ and show that their Poisson brackets correspond
to a double deformation of those of the \pcm. They correspond to
the Poisson brackets found in \cite{Balog:1993es} and to the ones
in \cite{Kawaguchi:2013gma} for the $\mathfrak{su}(2)$ case. We
show that the Poisson bracket of the Lax matrix \eqref{spatial lax}
takes the standard $r/s$-form \cite{Maillet:1985fn,Maillet:1985ek}
and identify the corresponding twist function \cite{Vicedo:2010qd}.

\subsection{Canonical analysis}

To perform the canonical analysis of the action \eqref{act-depart} we
introduce coordinates $\varphi^i$ on the group $G$ and write
\begin{equation*}
g^{-1} \partial_i g = L_i^A T_A,
\end{equation*}
in terms of a basis $\{ T_A \}$ of $\g$.
Letting $f_{AB}^{\phantom{AB}C}$ denote the structure constants with respect to this basis, namely $[T_A, T_B] = f_{AB}^{\phantom{AB}C} T_C$, we define
\begin{equation*}
\eta_{AB} = -\tr(T_A T_B), \qquad f_{ABC} = \eta_{CD} f_{AB}^{\phantom{AB}D}.
\end{equation*}
We also introduce the tensor $\lambda_{ij}=-\lambda_{ji}$ through the relation
\begin{equation*}
-\tr \Bigl( g^{-1} \partial_i g [ g^{-1} \partial_j g,g^{-1} \partial_k g]\Bigr)
=\partial_i \lambda_{jk} + \partial_j \lambda_{ki} +\partial_k \lambda_{ij} .
\end{equation*}
In terms of this, the WZ term in the action \eqref{act-depart} can be rewritten as
\begin{equation*}
S_{\rm WZ}[g] = k K \int d^2 x \partial_0 \varphi^i \partial_1 \varphi^j \lambda_{ij}.
\end{equation*}

The conjugate momenta $\pi_i$ of $\varphi^i$ can be computed from the action \eqref{act-depart} written in the form
\begin{align*}
S[g] & = - \ha K \int d^2 x \tr \Bigl(  g^{-1}\partial_0 g (1+ \eta^2 + \eta^2 R^2)  g^{-1}\partial_0 g
-  g^{-1}\partial_1 g (1+ \eta^2 + \eta^2 R^2)  g^{-1}\partial_1 g \\
&\qquad\qquad\qquad\qquad  + 2 g^{-1}\partial_0 g A R g^{-1}\partial_1 g
\Bigr)
+ k K \int d^2 x \partial_0 \varphi^i \partial_1 \varphi^j \lambda_{ij}.
\end{align*}
It is convenient to express the result in terms of the $\g$-valued field
\begin{equation*}
X = L_A^i \pi_i \eta^{AB} T_B,
\end{equation*}
where $L^i_A$ is defined as the inverse of $L_i^A$, namely $L^i_A L^B_i = \delta_A^B$.
Explicitly, we find that
\begin{equation} \label{X Igd0g}
X = K (1+ \eta^2 + \eta^2 R^2) g^{-1} \partial_0 g + K A R g^{-1} \partial_1 g + X_{\rm{WZ}}
\end{equation}
where
\begin{equation} \label{XWZ def}
X_{\rm WZ} = k K \lambda_{ij} \partial_1 \varphi^j L_A^i T^A.
\end{equation}
Note that using this last expression, the WZ term in the action \eqref{act-depart} can be written more succinctly as
\begin{equation*}
S_{\rm WZ}[g] = - \int d^2 x \tr(g^{-1} \partial_0 g X_{\rm WZ} ).
\end{equation*}

The canonical Poisson brackets between the coordinates $\varphi^i$ and their conjugate momenta $\pi_i$ may be conveniently expressed in terms of the fields $g$ and $X$ as
\begin{subequations} \label{g X XWZ PB}
\begin{align}
\{ g_{\1}, g_{\2} \} &= 0,\\
\{ X_{\1}, g_{\2} \} &= - g_{\2} C_{\1\2} \delta_{\sigma\sigma'},\\
\{ X_{\1} , X_{\2} \} &= [C_{\1\2}, X_{\2}] \delta_{\sigma\sigma'},
\end{align}
where to simplify the notation we use the convention that the argument of a function in the first (resp. second) tensor factor is $\sigma$ (resp. $\sigma'$). For instance,  $g_{\1} = g(\sigma) \otimes 1$ and $g_{\2} = 1 \otimes g(\sigma')$. The quadratic Casimir is $C_{\1\2}= \eta^{AB} T_A \otimes T_B$.
We also note the following relation
\begin{equation}
\{ X_{\1}, X_{\rm{WZ}\2} \} + \{ X_{\rm{WZ}\1}, X_{\2}  \}  = k K [C_{\1\2} ,
g_{\2}^{-1} \partial_{\sigma} g_{\2}]  \delta_{\sigma\sigma'} +
[C_{\1\2} , X_{\rm{WZ}\2}] \delta_{\sigma\sigma'}.
\end{equation}
\end{subequations}
In what follows it will be convenient to work with the $\g$-valued field $Y = X - X_{\rm WZ}$. Its Poisson brackets can be determined using \eqref{g X XWZ PB} and read
\begin{subequations} \label{g Y PB}
\begin{align}
  \{Y_{\1}, g_{\2} \} &= - g_{\2} C_{\1\2} \delta_{\sigma\sigma'},\\
\{ Y_{\1} , Y_{\2} \} &= [C_{\1\2} , Y_{\2} ] \delta_{\sigma\sigma'}  - k  K
 [C_{\1\2} , g_{\2}^{-1} \partial_{\sigma} g_{\2}] \delta_{\sigma\sigma'}.
\end{align}
\end{subequations}

Finally, the expression for the Hamiltonian is obtained as the Legendre transform of the Lagrangian. Explicitly we find
\begin{equation} \label{Ham}
H = - \frac{K}{2 \big( A^2 + (k-1)^2 \big) \big( A^2 + (k+1)^2 \big)} \int d\sigma \tr \big[ (1 + k^2 + A^2) (\KK_0^2 + \KK_1^2) + 4 k \KK_0 \KK_1 \big].
\end{equation}

\subsection{Current}

By using the relation \eqref{X Igd0g} we may express $g^{-1} \partial_0 g$
in terms of $g^{-1} \partial_1 g$ and $Y$. Substituting this into
the expressions \eqref{KP-KM} we find
\begin{subequations} \label{K0-K1}
\begin{align}
\KK_0 &= \frac{1}{K} g Y g^{-1} - k \partial_1 g g^{-1},\\
\KK_1 &= g\biggl( \frac{1}{K} \Bigl( - \frac{k}{1+\eta^2} + AR+
 \frac{ k\eta^2 }{1+\eta^2}R^2 \Bigr) Y + k \Bigl( \frac{1 + \eta^2}{k} + A R  + \frac{k \eta^2}{1+\eta^2} R^2 \Bigr) g^{-1} \partial_1 g \biggr) g^{-1}.
\end{align}
\end{subequations}
There are two interesting special limits of \eqref{K0-K1}. The first one is the
Yang-Baxter limit obtained by taking $k=0$, which implies $X_{\rm{WZ}}=0$.
By virtue of \eqref{A-k-eta} we then have $A=\eta$.
So in this limit the expressions \eqref{K0-K1} become
\begin{subequations} \label{K0-K1-YB}
\begin{align}
\KK_0 &= \frac{1}{K} g X   g^{-1} ,\\
\KK_1 &= \frac{\eta}{K} g RX g^{-1} + (1 + \eta^2  )  \partial_1 g g^{-1}.
\end{align}
\end{subequations}
This is in agreement with the expressions found in \cite{Delduc:2013fga} provided
we set $K=1+\eta^2$, which corresponds to the normalisation
of the action used there.
The second interesting limit corresponds to taking $\eta=0$ in which case $A=0$ and \eqref{K0-K1} reduce to
\begin{subequations} \label{K0-K1-WZ}
\begin{align}
\KK_0 &= \frac{1}{K} g Y g^{-1} - k \partial_1 g g^{-1},\\
\KK_1 &=  - \frac{k}{K} g Y g^{-1} +  \partial_1 g g^{-1}.
\end{align}
\end{subequations}
Finally, the $G_L \times G_R$ invariance of the \pcm is broken down to $G_L \times H_R$ by the deformation, where $H$ is the Cartan subgroup of $G$. Note that $\int d\sigma \KK_0$ is the charge which generates the unbroken symmetry $G_L$. This preservation of the $G_L$ symmetry is in contrast with the situation in the Bi-Yang-Baxter $\sigma$-model \cite{Klimcik:2014bta}, where both $G_L$ and $G_R$ symmetries are broken.

\subsection{Two-parameter deformed Poisson brackets}

Given the expressions \eqref{K0-K1} for the fields $\KK_0$, $\KK_1$ in terms
of $g$ and $Y$, we may compute their Poisson brackets using \eqref{g Y PB}.
After a direct but lengthy calculation using the modified classical
Yang-Baxter equation \eqref{YB R} for $R$, its consequence
\eqref{YB R2} for $R^2$ and the expression \eqref{A-k-eta} for $A$
together with the identities \eqref{id R2} and
\begin{equation*}
(1 + R^2) \big[ x, (1 + R^2)y \big] = 0,
\end{equation*}
valid for any $x, y \in \g$, we find
\begin{subequations} \label{collect-pb-k}
\begin{align}
\{  \KK_{0\1}, \KK_{0\2} \} &=  -\frac{1}{K} [C_{\1\2}, \KK_{0\2}] \dss -\frac{2k}{K} C_{\1\2} \pdss,\\
\{ \KK_{0\1}, \KK_{1\2} \} &=  -\frac{1}{K} [C_{\1\2}, \KK_{1\2}] \dss +
 \frac{1 + k^2 + A^2}{K}  C_{\1\2} \pdss,\\
 \{ \KK_{1\1}, \KK_{1\2} \} &= \frac{k^2 + A^2}{K} [C_{\1\2}, \KK_{0\2}] \dss
+ \frac{2k}{K} [C_{\1\2}, \KK_{1\2}] \dss - \frac{2k}{K} C_{\1\2} \pdss.
\end{align}
\end{subequations}
In the $\mathfrak{su}(2)$ case, this Poisson bracket is exactly
the one of the `squashed WZW model' which was identified
in \cite{Kawaguchi:2013gma}, up to the overall factor of $1/K$.
The bracket \eqref{collect-pb-k} was also studied more recently
in \cite{Itsios:2014vfa} for a general Lie algebra. The notation
used there for the current components $\mathcal{I}_0$, $\mathcal{I}_1$,
and for the parameters $\rho$, $x$ and $e^2$ may be identified with
the present notation as
\begin{gather*}
\KK_0 = - \mathcal{I}_0, \quad
\KK_1 = - \mathcal{I}_1, \quad
K = \frac{1}{2 e^2} \frac{1}{1 + \rho^2 + x (1 - \rho^2)}, \\
k = 2 e^2 K \rho, \quad
A^2 = 4 e^4 K^2 (1 - x^2) (1 - \rho^2)^2.
\end{gather*}
In particular, since $A$ must be a real parameter in the action
\eqref{act-depart}, we have $A^2 \geq 0$ and so we see that the
Poisson brackets \eqref{collect-pb-k} correspond to $-1 \leq x \leq 1$,
which is known to correspond to the complex branch \cite{Hollowood:2014rla}.

Let us consider the $k=0$ limit of these Poisson brackets. If we
choose the normalisation of the action to be $K= 1+ \eta^2$ then we find
\begin{subequations} \label{pb-k-yb}
\begin{align}
\{  \KK_{0\1}, \KK_{0\2} \} &=  -\frac{1}{1+\eta^2} [C_{\1\2}, \KK_{0\2}] \dss ,\\
\{ \KK_{0\1}, \KK_{1\2} \} &=  -\frac{1}{1+\eta^2} [C_{\1\2}, \KK_{1\2}] \dss +
   C_{\1\2} \pdss,\\
 \{ \KK_{1\1}, \KK_{1\2} \} &= \frac{\eta^2}{1 + \eta^2} [C_{\1\2}, \KK_{0\2}] \dss.
\end{align}
\end{subequations}
This agrees with the one-parameter deformation of the Poisson brackets of the principal chiral model (see for instance \cite{Delduc:2013fga}).

\subsection{Twist function}

The Poisson brackets \eqref{collect-pb-k} can be written in terms of the Lax matrix \eqref{spatial lax} as
\begin{equation} \label{Lax algebra}
\{ \L_{\1}(z), \L_{\2}(z') \} = [r_{\1\2}, \L_{\1}(z) + \L_{\2}(z')] \delta_{\sigma, \sigma'} - [s_{\1\2}, \L_{\1}(z) - \L_{\2}(z')] \delta_{\sigma, \sigma'} - 2 s_{\1\2} \delta'_{\sigma \sigma'},
\end{equation}
where the $r/s$-matrices read
\begin{equation*}
r_{\1\2}(z, z') = \frac{\varphi_{\eta, k}(z)^{-1} + \varphi_{\eta, k}(z')^{-1}}{z - z'} C_{\1\2}, \qquad
s_{\1\2}(z, z') = \frac{\varphi_{\eta, k}(z)^{-1} - \varphi_{\eta, k}(z')^{-1}}{z - z'} C_{\1\2},
\end{equation*}
and the deformed twist function $\varphi_{\eta, k}(z)$ is given by
\begin{equation} \label{twist func}
\varphi_{\eta, k}(z) = \frac{K (1 - z^2)}{A^2 + (z - k)^2}.
\end{equation}
Here $A$ is defined as in \eqref{A-k-eta}. As usual, the form \eqref{Lax algebra} for the Poisson bracket of the Lax matrix implies the existence of an infinite number of Poisson commuting quantities \cite{Maillet:1985fn,Maillet:1985ek}.  
Finally, let us note that, just as in the case of the 1-parameter deformed Poisson brackets \cite{Rajeev:1988hq}, one can also recast the Poisson bracket \eqref{collect-pb-k} in the form of two Poisson commuting complex Kac-Moody algebras.
In general, Kac-Moody currents can be constructed by taking the Lax matrix at the simple poles of the twist function \cite{del-inprep}. In the present case,
if we define the currents
\begin{equation*}
\mathscr J_{\pm} = \L(k \pm i A) = \frac{1}{1 - (k \pm i A)^2} \big( \KK_1 + (k \pm i A) \KK_0 \big),
\end{equation*}
then ${\mathscr J}^{\dag}_- = \mathscr J_+$ and these satisfy the following Poisson brackets
\begin{align*}
\gamma_{\pm} \big\{ \mathscr J_{\pm\1}, \mathscr J_{\pm\2} \big\} &= - \big[ C_{\1\2}, \mathscr J_{\pm\2} \big] \delta_{\sigma \sigma'} + C_{\1\2} \delta'_{\sigma \sigma'},\\
\big\{ \mathscr J_{+\1}, \mathscr J_{-\2} \big\} &= 0,
\end{align*}
where $\gamma_{\pm} = \pm \frac{K}{2 i A} (1 - (k \pm i A)^2)$.

\section{Conclusion}

In this note we presented a two-parameter deformation of the principal chiral model.

Let us emphasise that our proof of the integrability of the model defined by the action \eqref{act-depart} makes use of the identity \eqref{R3 -R} for the $R$-matrix. Although the latter is certainly satisfied by the standard $R$-matrix, it doesn't hold in general. The more general case deserves further study.

The deformed model furnishes a higher rank generalisation of the `squashed WZNW-model' introduced in \cite{Kawaguchi:2011mz,Kawaguchi:2013gma}. The action of the latter is given simply by adding a Wess-Zumino term to the squashed 3-sphere $\sigma$-model action. A peculiarity of the $\mathfrak{su}(2)$ case, however, is the absence of a $B$-field in the Yang-Baxter $\sigma$-model. In the higher rank cases, where a $B$-field is present, we see that the action \eqref{act-depart} is not obtained from the Yang-Baxter $\sigma$-model action by the mere addition of  a Wess-Zumino term. Indeed, the relative weight of the metric and $B$-field terms in the Yang-Baxter $\sigma$-model action also has to be suitably deformed.

As already emphasised, the model  provides a realisation of the two-parameter family of deformed Poisson brackets \cite{Balog:1993es} in the complex branch.
It would be very interesting to identify the integrable $\sigma$-model realising the deformed Poisson bracket in the real branch. Although the latter is not known, it was recently shown in \cite{Itsios:2014vfa} that this hypothetical model admits a classical Yangian symmetry $Y_C(\g)$. It would be interesting to identify the full symmetry algebra of the model in the complex branch as well, using the methods developed in \cite{Kawaguchi:2011pf,Kawaguchi:2012gp,Delduc:2013fga,Kawaguchi:2011mz,Kawaguchi:2012ve}.

\paragraph{Acknowledgements.} We thank B. Hoare for useful discussions. This work is partially supported by the program PICS 6412 DIGEST of CNRS.

\providecommand{\href}[2]{#2}\begingroup\raggedright\endgroup


\begin{thebibliography}{10}

\bibitem{Rajeev:1988hq}
S.~Rajeev, {\it {Nonabelian bosonization without Wess-Zumino terms. I. New
  current algebra}},  {\em Phys. Lett.} {\bf B217} (1989) 123.

\bibitem{Hollowood:2014rla}
T.~J. Hollowood, J.~L. Miramontes, and D.~M. Schmidtt, {\it {Integrable
  Deformations of Strings on Symmetric Spaces}},
  \href{http://xxx.lanl.gov/abs/1407.2840}{{\tt arXiv:1407.2840}}.

\bibitem{Delduc:2013fga}
F.~Delduc, M.~Magro, and B.~Vicedo, {\it {On classical $q$-deformations of
  integrable $\sigma$-models}},  {\em JHEP} {\bf 1311} (2013) 192,
  [\href{http://xxx.lanl.gov/abs/1308.3581}{{\tt arXiv:1308.3581}}].

\bibitem{Klimcik:2002zj}
C.~Klimcik, {\it {Yang-Baxter sigma models and dS/AdS T duality}},  {\em JHEP}
  {\bf 0212} (2002) 051, [\href{http://xxx.lanl.gov/abs/hep-th/0210095}{{\tt
  hep-th/0210095}}].

\bibitem{Klimcik:2008eq}
C.~Klimcik, {\it {On integrability of the Yang-Baxter $\sigma$-model}},  {\em
  J. Math. Phys.} {\bf 50} (2009) 043508,
  [\href{http://xxx.lanl.gov/abs/0802.3518}{{\tt arXiv:0802.3518}}].

\bibitem{Kawaguchi:2011pf}
I.~Kawaguchi and K.~Yoshida, {\it {Hybrid classical integrability in squashed
  sigma models}},  {\em Phys. Lett.} {\bf B705} (2011) 251--254,
  [\href{http://xxx.lanl.gov/abs/1107.3662}{{\tt arXiv:1107.3662}}].

\bibitem{Kawaguchi:2012gp}
I.~Kawaguchi, T.~Matsumoto, and K.~Yoshida, {\it {On the classical equivalence
  of monodromy matrices in squashed sigma model}},  {\em JHEP} {\bf 1206}
  (2012) 082, [\href{http://xxx.lanl.gov/abs/1203.3400}{{\tt
  arXiv:1203.3400}}].

\bibitem{Sfetsos:2013wia}
K.~Sfetsos, {\it {Integrable interpolations: From exact CFTs to non-Abelian
  T-duals}},  {\em Nucl. Phys.} {\bf B880} (2014) 225--246,
  [\href{http://xxx.lanl.gov/abs/1312.4560}{{\tt arXiv:1312.4560}}].

\bibitem{Itsios:2014vfa}
G.~Itsios, K.~Sfetsos, K.~Siampos, and A.~Torrielli, {\it {The classical
  Yang-Baxter equation and the associated Yangian symmetry of gauged WZW-type
  theories}},  {\em Nucl. Phys.} {\bf B889} (2014) 64--86,
  [\href{http://xxx.lanl.gov/abs/1409.0554}{{\tt arXiv:1409.0554}}].

\bibitem{Sfetsos:2014cea}
K.~Sfetsos and D.~C. Thompson, {\it Spacetimes for $\lambda$-deformations},
  \href{http://xxx.lanl.gov/abs/1410.1886}{{\tt arXiv:1410.1886}}.

\bibitem{Balog:1993es}
J.~Balog, P.~Forgacs, Z.~Horvath, and L.~Palla, {\it {A New family of SU(2)
  symmetric integrable sigma models}},  {\em Phys. Lett.} {\bf B324} (1994)
  403--408, [\href{http://xxx.lanl.gov/abs/hep-th/9307030}{{\tt
  hep-th/9307030}}].

\bibitem{Kawaguchi:2011mz}
I.~Kawaguchi, D.~Orlando, and K.~Yoshida, {\it {Yangian symmetry in deformed
  WZNW models on squashed spheres}},  {\em Phys. Lett.} {\bf B701} (2011)
  475--480, [\href{http://xxx.lanl.gov/abs/1104.0738}{{\tt arXiv:1104.0738}}].

\bibitem{Kawaguchi:2013gma}
I.~Kawaguchi and K.~Yoshida, {\it {A deformation of quantum affine algebra in
  squashed Wess-Zumino-Novikov-Witten models}},  {\em J. Math. Phys.} {\bf 55}
  (2014) 062302, [\href{http://xxx.lanl.gov/abs/1311.4696}{{\tt
  arXiv:1311.4696}}].

\bibitem{Maillet:1985fn}
J.~M. Maillet, {\it {Kac-Moody algebra and extended Yang-Baxter relations in
  the $O(N)$ non-linear sigma model}},  {\em Phys. Lett.} {\bf B162} (1985)
  137.

\bibitem{Maillet:1985ek}
J.~M. Maillet, {\it {New integrable canonical structures in two-dimensional
  models}},  {\em Nucl. Phys.} {\bf B269} (1986) 54.

\bibitem{Vicedo:2010qd}
B.~Vicedo, {\it {The classical R-matrix of AdS/CFT and its Lie dialgebra
  structure}},  {\em Lett. Math. Phys.} {\bf 95} (2011) 249--274,
  [\href{http://xxx.lanl.gov/abs/1003.1192}{{\tt arXiv:1003.1192}}].

\bibitem{Klimcik:2014bta}
C.~Klimcik, {\it {Integrability of the Bi-Yang-Baxter $\sigma$-model}},  {\em
  Lett. Math. Phys.} {\bf 104} (2014) 1095--1106,
  [\href{http://xxx.lanl.gov/abs/1402.2105}{{\tt arXiv:1402.2105}}].

\bibitem{del-inprep}
F.~Delduc, M.~Magro, and B.~Vicedo,  In preparation.

\bibitem{Kawaguchi:2012ve}
I.~Kawaguchi, T.~Matsumoto, and K.~Yoshida, {\it {The classical origin of
  quantum affine algebra in squashed sigma models}},  {\em JHEP} {\bf 1204}
  (2012) 115, [\href{http://xxx.lanl.gov/abs/1201.3058}{{\tt
  arXiv:1201.3058}}].

\end{thebibliography}
\end{document}